\documentclass[10pt]{scrartcl}
\usepackage{amsmath,amssymb}
\usepackage{amsthm}
\usepackage{slashed}
\usepackage[pdftex]{graphicx}
\usepackage[headsep=15pt,head=80pt,foot=20pt,margin=60pt]{geometry}
\usepackage{arev}

\def\heart{\lower.1em\hbox{$\heartsuit$}}
\let\tilde\widetilde

\newcommand{\Group}[2]{{ \hbox{{\itshape{#1}}($#2$)} }}
\newcommand{\U}[1]{\Group{U\kern0.05em}{#1}}
\newcommand{\SU}[1]{\Group{SU\kern0.1em}{#1}}
\newcommand{\SL}[1]{\Group{SL\kern0.05em}{#1}}
\newcommand{\Sp}[1]{\Group{Sp\kern0.05em}{#1}}
\newcommand{\SO}[1]{\Group{SO\kern0.1em}{#1}}

\newcommand{\mybar}[1]%
    {{\kern 0.8pt\overline{\kern -0.8pt#1\kern -0.8pt}\kern 0.8pt}}
\newcommand{\sla}[1]%
    {{\raise.15ex\hbox{$/$}\kern-.57em #1}}
\newcommand{\roughly}[1]%
    {{ \mathrel{\raise.3ex\hbox{ $#1$\kern-.75em\lower1ex\hbox{$\sim$}} } }}

\newcommand{\nop}[1]{:\kern-.3em#1\kern-.3em:}

\newcommand{\delfb}{\overleftrightarrow{\partial}}

\newcommand{\al}{\ensuremath{\alpha}}

\newcommand{\ga}{\ensuremath{\gamma}}

\newcommand{\et}{\ensuremath{\eta}}

\newcommand{\si}{\ensuremath{\sigma}}

\newcommand{\ph}{\ensuremath{\phi}}

\newcommand{\n}{\notag \\}
\newcommand{\mcl}[1]{\mathcal{#1}}

\numberwithin{equation}{section}
\numberwithin{figure}{section}

\begin{document}
\thispagestyle{empty}

\begin{titlepage}

\begin{flushright}
UG-FT-312/14\\
CAFPE-182/14 
\end{flushright}

\vskip 5em

\begin{center}
\textbf{\Large 
 Unitarity bounds on scalar dark matter effective interactions at LHC\\
}

\vskip 4em

Yasuhiro Yamamoto

\vskip 4em

\textit{
  Deportamento de Fisica Teorica y del Cosmos,\\
  Facultad de Ciencias, Universidad de Granada,
  Granada E-18071 Spain}\\

\vskip 4em

\textbf{Abstract}
\end{center}

\medskip
\noindent
We study the compatibility of the unitarity bound and the 8TeV LHC on the effective theory of the scalar dark matter.
In several signals of effective interactions, mono-jet with missing energy events are studied.
We found that, at least, if the dark matter mass is about 800GeV or heavier, contributions of events violating the unitarity are not negligible.
The unitarity conditions in the 14TeV LHC are also calculated.

\bigskip
\vfill

\end{titlepage}

\section{Introduction}
Astrophysical observations have clarified enormous amount of dark matter exist in our universe.
A promising scenario is that the dark matter is a new particle weakly interacting with the Standard Model (SM) particles.
We, however, have not found any clue of it.

Collider experiments are an important tool to shed light on it, where it is detected as excess of SM particles with large missing energy events.
If the dark matter exit in the reach of colliders, which is expected by the WIMP miracle, we could find its various information, for instance, the mass, the spin, couplings with the SM, and even something about mediators.
Since interactions between the dark matters and the SM particles should be very weak, analyses with the effective field theory has been used well, where we can obtain new information of the dark matter without assuming a certain UV structure~\cite{Beltran:2010ww}.

ATLAS~\cite{RefATLAS} and CMS~\cite{RefCMS} have submitted bounds to effective interactions between the dark matters and colored particles.
They have investigated the monojet with missing energy events.
Dominant contributions of the process are that, after emitting a jet, collisions of two partons produce the dark matter pairs.
If the dark matter is heavy, large energy is required for the pair creation.
In other words, large energy is injected into the effective vertex.
This means that these events are simultaneously in danger of violating the unitarity bound~\cite{Jacob:1959at}.

If the number of events violating the bounds is not small, given limits to the effective interactions are not reliable.
People recently have studied it using several explicit simple UV completions\cite{Busoni:2013lha}.
However, compatibility of these two issues have not been seriously investigated until a recent work~\cite{RefUnitarity}.
In the work, the dark matter is assumed to be a fermion.
Properties of effective interactions are different if the dark matter is a scalar field.
Hence, we discuss relations between experimental bounds and the unitarity conditions of the scalar dark matter with the effective field theory.

The rest of this paper is organized as follows.
In the next section, Sec.~\ref{SecEffective}, we introduce effective interactions studied in this paper.
The importance of operator dimension is also pointed out there.
Obtained conditions are applied to collider studies, and they are compared with current experimental results in Sec.~\ref{SecUnitarity}.
Consequences of our studies are summarized in Sec.~\ref{SecConclusion}.
The appendix is devoted to show more general formulae of the unitarity bounds for the scalar dark matters.
\section{Effective interactions and unitarity bounds}
\label{SecEffective}
We study the following three effective interactions~\footnotemark,
\begin{align}
 \text{the pseudo scalar interaction:}& \quad
  \frac{i}{M_P} \ph^\dag \ph (\bar{q}\ga_5 q), \\
 \text{the axial vector interaction:}& \quad
  \frac{i}{M_A^2} (\ph^\dag \delfb_\mu \ph) (\bar{q} \ga^\mu \ga_5 q), \\
 \text{the pseudo gluon interaction:}& \quad
  \frac{g_s^2}{(8\pi)^2 M_\text{CS}^2} \ph^\dag \ph G^{a\mu\nu} \tilde{G}^a{}_{\mu\nu}.
\end{align}
In the above interactions, $\ph$ is the complex scalar dark matter, $q$ stands for quarks, $G^a_{\mu\nu}$ is the field strength tensor of gluon, and $g_S$ is the coupling of the SM $SU(3)_c$.
\footnotetext{
 Since they are the spin dependent interactions, direct detection bounds can be evaded without tunings like the isospin violation.
 Several constraints to these interactions have been studied in \cite{Bhattacherjee:2012ch}.
 How to obtain UV completions generating them is discussed in \cite{Busoni:2013lha,Chang:2013oia}.
}
If the operators satisfy the thermal abundance, i.e. $\langle \si v_\text{rel} \rangle \sim 0.1$pb, the above suppression scales are
\begin{align}
 M_P \sim & 130\, \text{[TeV]},\\
 M_A \sim & 2.6 \sqrt{\frac{m_\text{DM}}{1\text{TeV}} } \, \text{[TeV]},\\
 M_{CS} \sim & 1.2 \sqrt{\frac{m_\text{DM}}{1\text{TeV}} } \, \text{[TeV]}.
\end{align}

Following the discussion in Ref.~\cite{RefUnitarity}, these suppression scales of the above coefficients are restricted by the $S$-matrix unitarity of parton level $\bar{q}q$ or $gg \to \ph^\dag \ph$ subprocesses,
\begin{align}
 M_\text{P} & \geq
   \frac{\sqrt{s}}{8\pi} \left( 1-\frac{4m_\text{DM}^2}{s} \right)^{1/4}, \\
 M_\text{A} & \geq 
   \frac{1}{2} \sqrt{\frac{s}{3\sqrt{2}\pi}} \left( 1-\frac{4m_\text{DM}^2}{s} \right)^{3/8}, \\
 M_\text{CS} & \geq 
   \frac{g_s}{16\pi} \sqrt{\frac{2s}{\pi}} \left( 1-\frac{4m_\text{DM}^2}{s} \right)^{1/8}.
 \label{EqGluon}
\end{align}
In these expressions, $\sqrt{s}$ is the invariant mass of the produced DM pair, $m_{DM}$ is the mass of the dark matter.
Some of other effective interactions also appear in this order.
We do not care them in our numerical studies below since they are similar to one of our following results.
Unitarity conditions including them are shown in the appendix.
Differences from the real scalar dark matter are also discussed there.

Qualitative consequences of the above conditions are different depending on their operator dimensions.
Let us consider a effective interaction,
\begin{align}
 \mcl{L} = \frac{1}{M^{D-4}} O_D,
\end{align}
where $O_D$ is a dimension $D$ operator.
The cross section of the operator can be simply written like,
\begin{align}
 \si \propto \frac{s^{D-5}}{M^{2(D-4)}}.
\end{align}
This cross section rapidly grows as increasing of collision energy if the operator dimension is higher than five.
Hence, the restriction by the unitarity become stronger if the dimensions of operators are higher.
We study a dimension five operator and two dimension six operators.
These differences of behaviors are numerically shown in the next section~\footnotemark.
\footnotetext{
 This feature is also observed in Ref.~\cite{RefUnitarity}, where a dimension seven operator has been studied.
}
\section{Unitarity bounds at LHC experiments}
\label{SecUnitarity}
The monojet searches in the 8TeV LHC were studied in ATLAS~\cite{RefATLAS} and CMS~\cite{RefCMS}.
Any excess has not been observed yet, then they have obtained lower bounds of the suppression scales.
We follow the analysis of CMS because their luminosity used in the analysis is about twice larger than the one used by ATLAS.

We made a model file including the effective interactions with FeynRules~\cite{Alloul:2013bka}, and generated monojet events for each interaction with MadGraph5~\cite{Madgraph}, where CTEQ parton distribution functions~\cite{Pumplin:2002vw} are used.
For simplicity, we have analysed events at the parton level.
Various transverse momentum cuts and the pseudo rapidity cut, $|\et|\leq 2.4$, have been applied to the visible particle in each final state.
In order to correct differences from the full simulation, $p\, p \to Z (\to \nu \bar{\nu})\, j$ has also been generated, and it is compared with the simulations of the SM background in Ref.~\cite{RefCMS}.
The strongest experimental bounds to the suppression scales are obtained when the $p_T$ cut is 450GeV~\footnotemark.
\footnotetext{
 In this case, the parton level cross section of $p\, p \to Z (\to \nu \bar{\nu})\, j$ is 63.3fb, and the given event number is 1247 with 19.7fb$^{-1}$, while the full simulation result is 1460.
 Then, the correction factor is 1.17.
}
Estimated lower limits are shown as blue/dark grey curves in figures below.

We checked whether each event satisfies the unitarity condition or not, and counted the number of events.
Satisfying the condition does not mean the reliability of results.
Events near the unitarity bound could not reproduce correct results as an effective theory even if they do not violate the unitarity.
However, if events violating the condition are not small rate in a process, it is clear that given results are not reliable anymore.

In monojet events, one of colored particle in the effective vertices must be virtual, and sub diagrams with the dark matters are not $2\to2$ in some of diagrams.
Therefore, several events are not correctly treated in the above prescription.
Their contaminations are, however, small as discussed in Ref.~\cite{RefUnitarity}.
We do not take care them in the following analyses.

Constraints from the unitarity bounds are also studied in the 14TeV run of LHC.
Here, we have assumed that cut conditions are the same with the 8TeV run.
A few years later, we should change these results putting actual bounds of the 14TeV run if the dark matter has not been detected.

\subsection{Pseudo scalar interaction}
Firstly, we study the pseudo scalar interaction between the complex scalar dark matter and quarks.
This is a dimension five operator.
The parton level cross section of this operator is almost independent of collision energy, as mentioned in the previous section.
Then, contributions of low energy scatterings become relatively large because of large parton luminosity, so that, the unitarity condition does not strongly restrict experimental results.

\begin{figure}[t]
\centering
 \includegraphics[scale=0.4,clip]{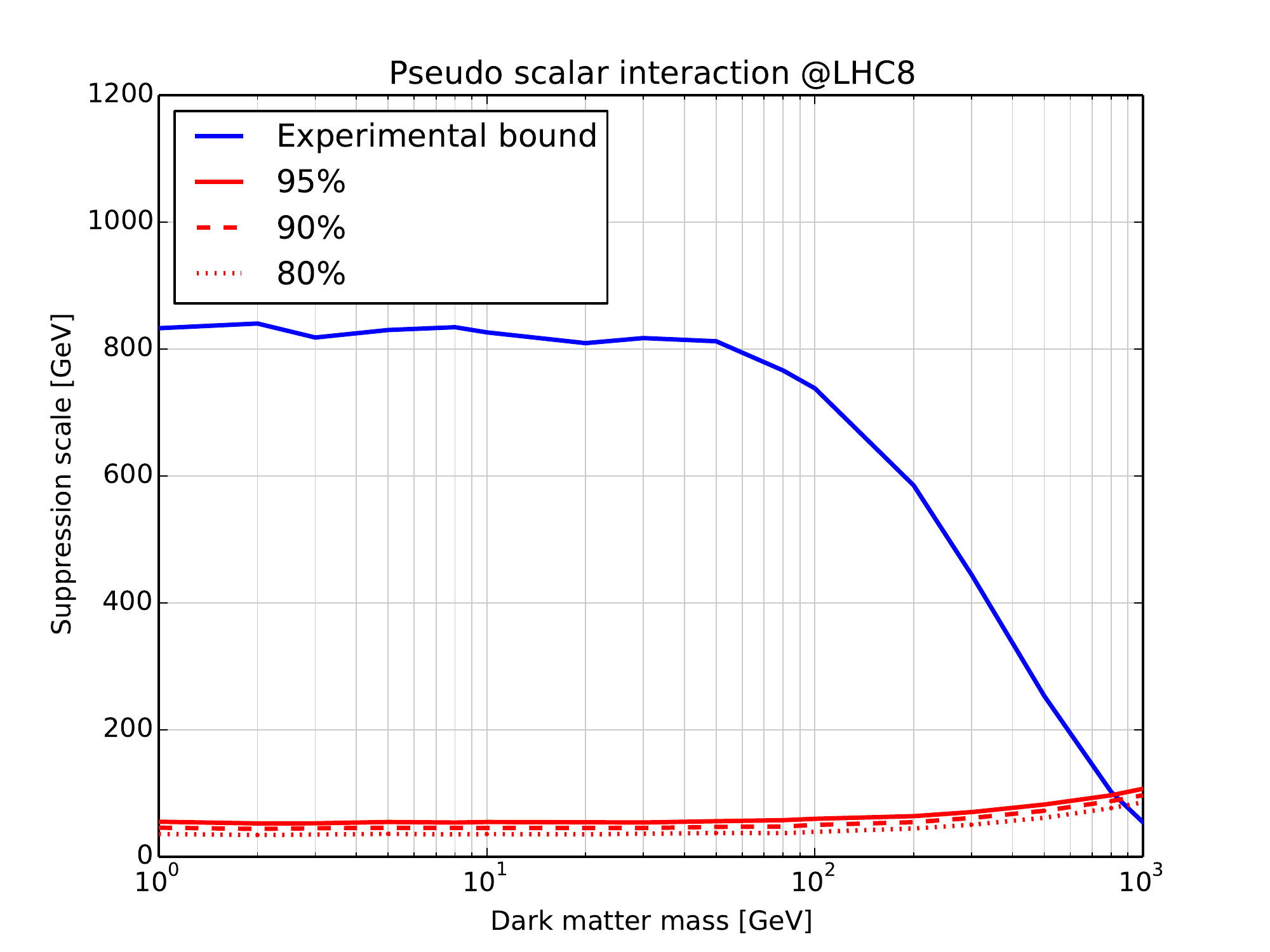}
 \includegraphics[scale=0.4,clip]{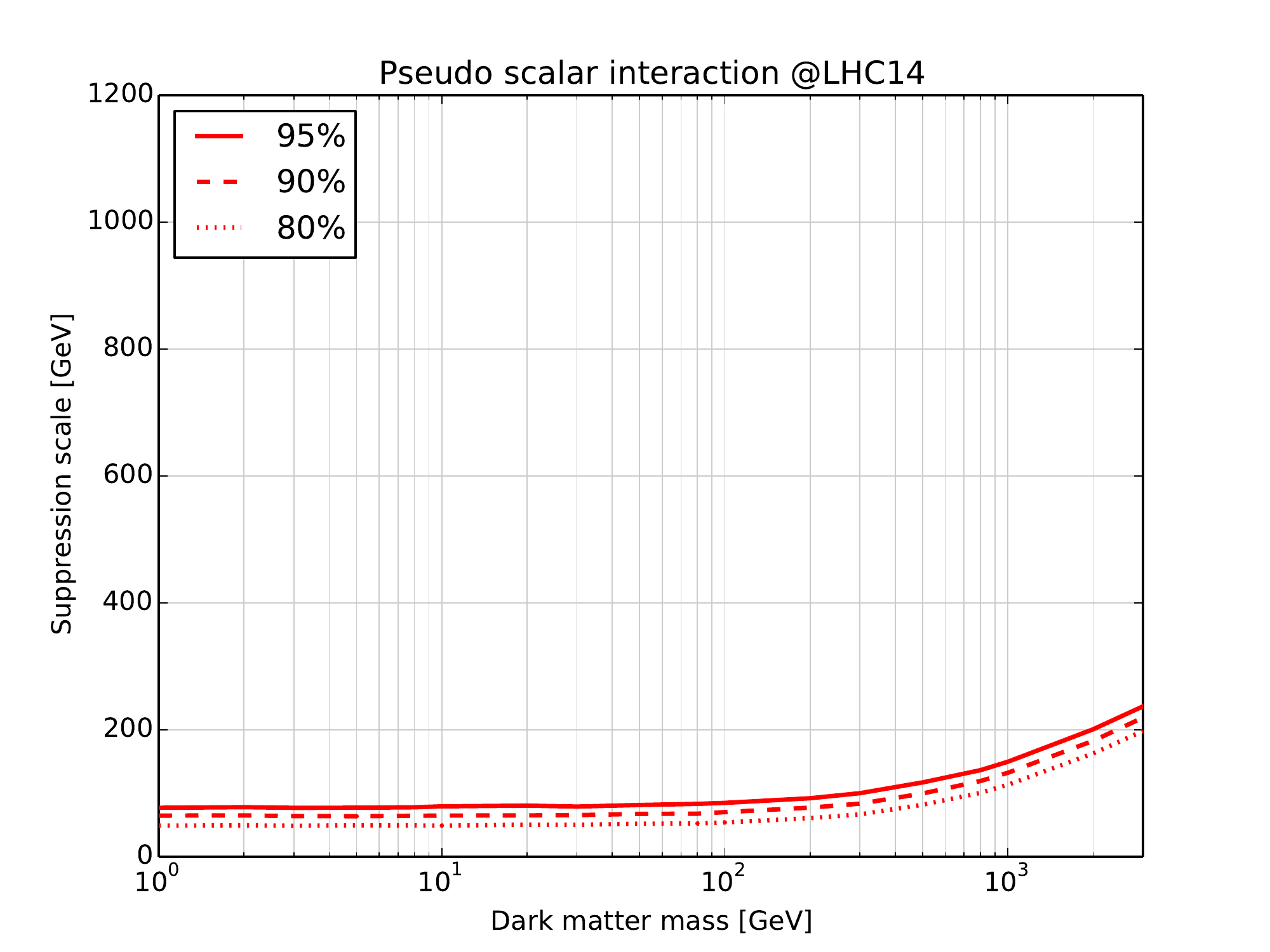}
\caption{
  The unitarity violation rate of the pseudo scalar interaction in LHC8(left) and LHC14(right).
	The vertical and the horizontal lines respectively stand for the suppression scales and the dark matter masses.
	The blue/dark grey line is the lower bound of the suppression scale calculated with experimental results in Ref.~\cite{RefCMS}.
	For each dark matter mass, 95\%, 90\% and 80\% of events satisfy the unitarity condition at the red/light grey solid, broken and dotted lines, respectively.}
\label{FigScalar}
\end{figure}

The experimental bound and the unitarity constraints are shown in Fig.~\ref{FigScalar}.
Because of the flat cross section, the red/light grey lines, which represent the event rates violating the unitarity, almost degenerate.
According to the figure, monojet analysis with this interaction is not valid if the dark matter is heavier than about 800GeV.
The bound evaluated with the experimental result of this higher dimensional operator is not contaminated by the unitarity condition even if the given suppression scale is about 200GeV.

In the 14TeV run, the condition is not so largely changed comparing to that in the 8TeV run because of its operator dimension.

\subsection{Axial vector interaction}
The second one is the axial vector interaction.
The dimension of this operator is six, so that, the results are qualitatively different from the above one.
The parton level cross section of this operator is proportional to $s/M^4$.
To produce heavy dark matters, large collision energy is required.
These events are enhanced by that factor, and simultaneously suffered by the constraint of the unitarity.

\begin{figure}[t]
\centering
 \includegraphics[scale=0.4,clip]{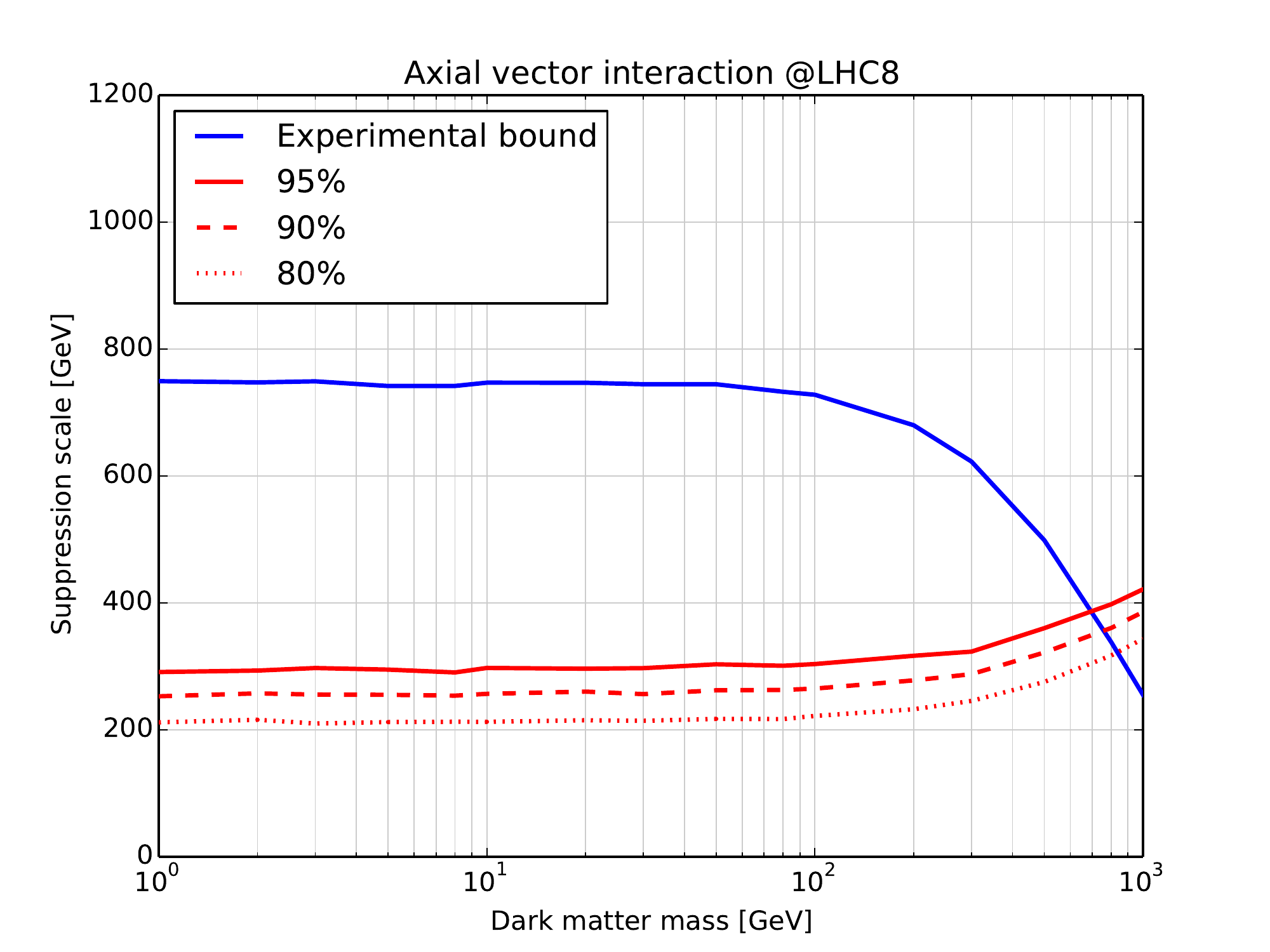}
 \includegraphics[scale=0.4,clip]{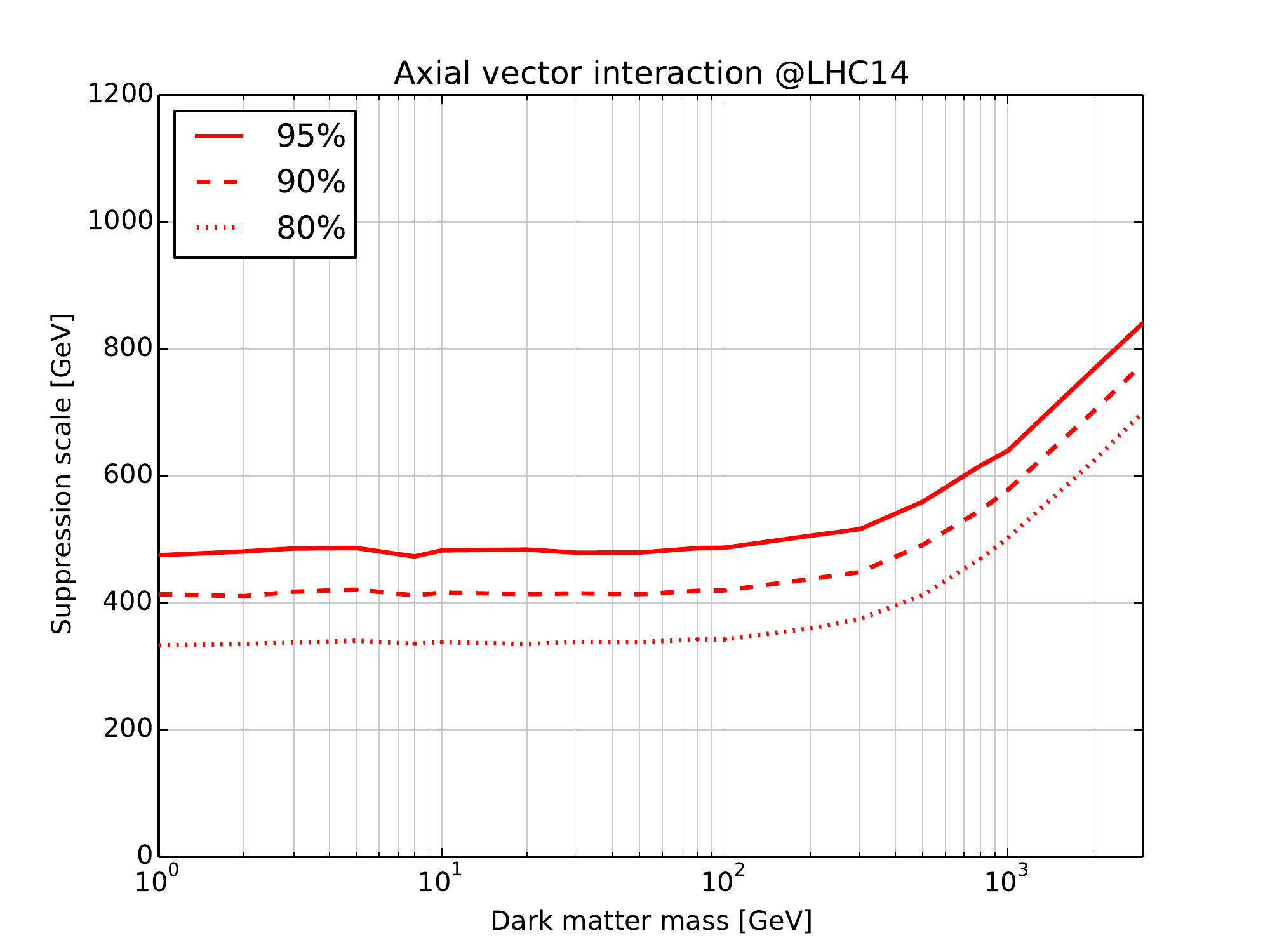}
\caption{The unitarity violation rate of the axial vector interaction in LHC8(left) and LHC14(right). The meanings of the axes and lines are the same with Fig.~\ref{FigScalar}}
\label{FigVector}
\end{figure}

Fig.~\ref{FigVector} states 10\% of events violate the unitarity if the dark matter mass is about 800GeV.
This result is numerically similar to the scalar interaction.
This is because the enhancement keeps large cross sections even in heavy dark matter region, while the unitarity conditions become stronger.
Hence, the experimental bound to the scale at the point is about 400GeV, which is much higher than the one for the scalar interaction.

In the 14TeV run, contributions of higher energy scatterings become larger, so that, the unitarity condition more strongly restricts the suppression scale.
For instance, at $m_{DM} = 1$TeV, 10\% of events violate the condition when the suppression scale is about 600GeV in LHC14, which is 400GeV in LHC8.
The lines of the unitarity steeply rise in the TeV region.

\subsection{Pseudo gluon interaction}
\begin{figure}[t]
 \includegraphics[scale=0.4,clip]{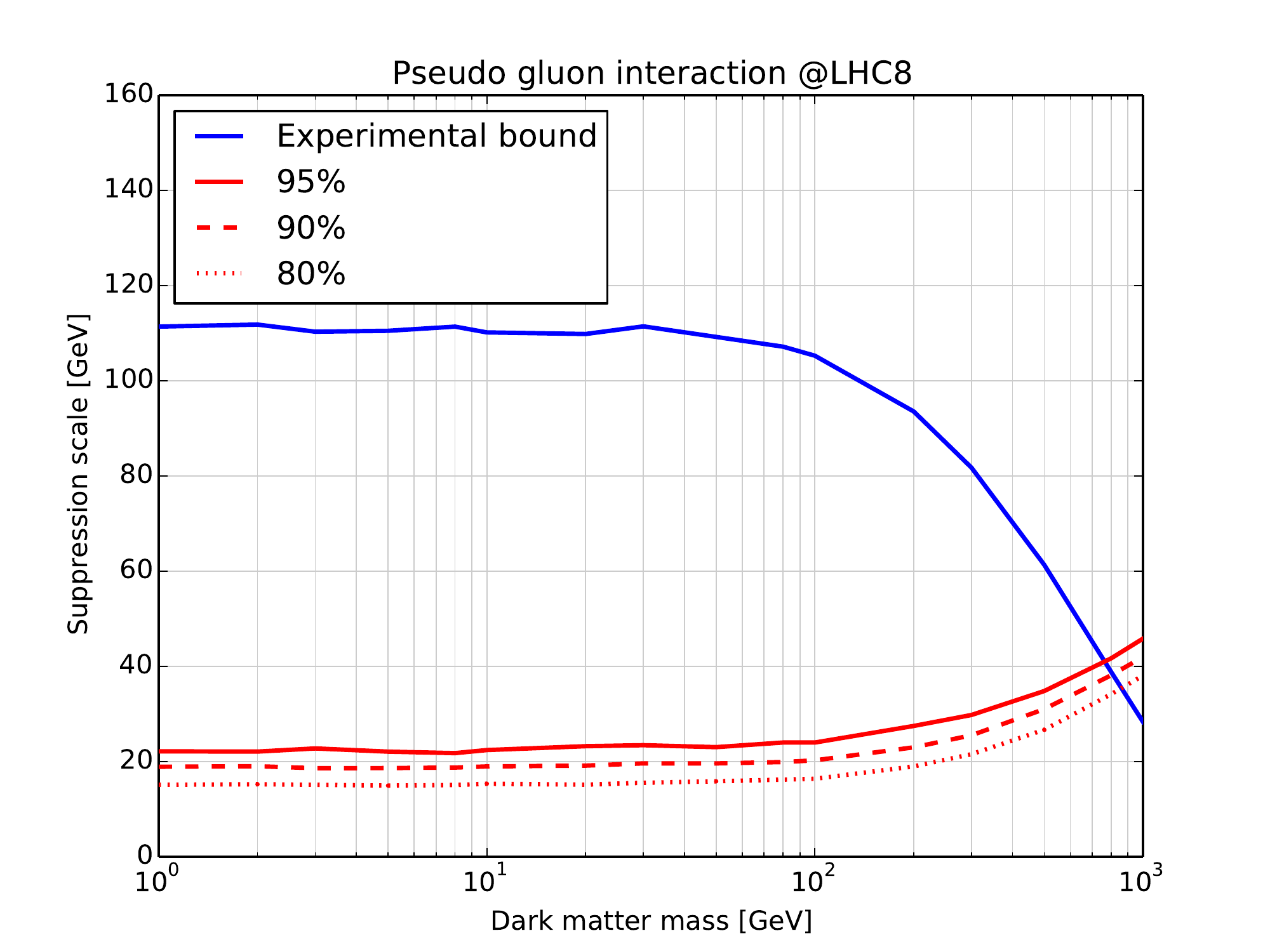}
 \includegraphics[scale=0.4,clip]{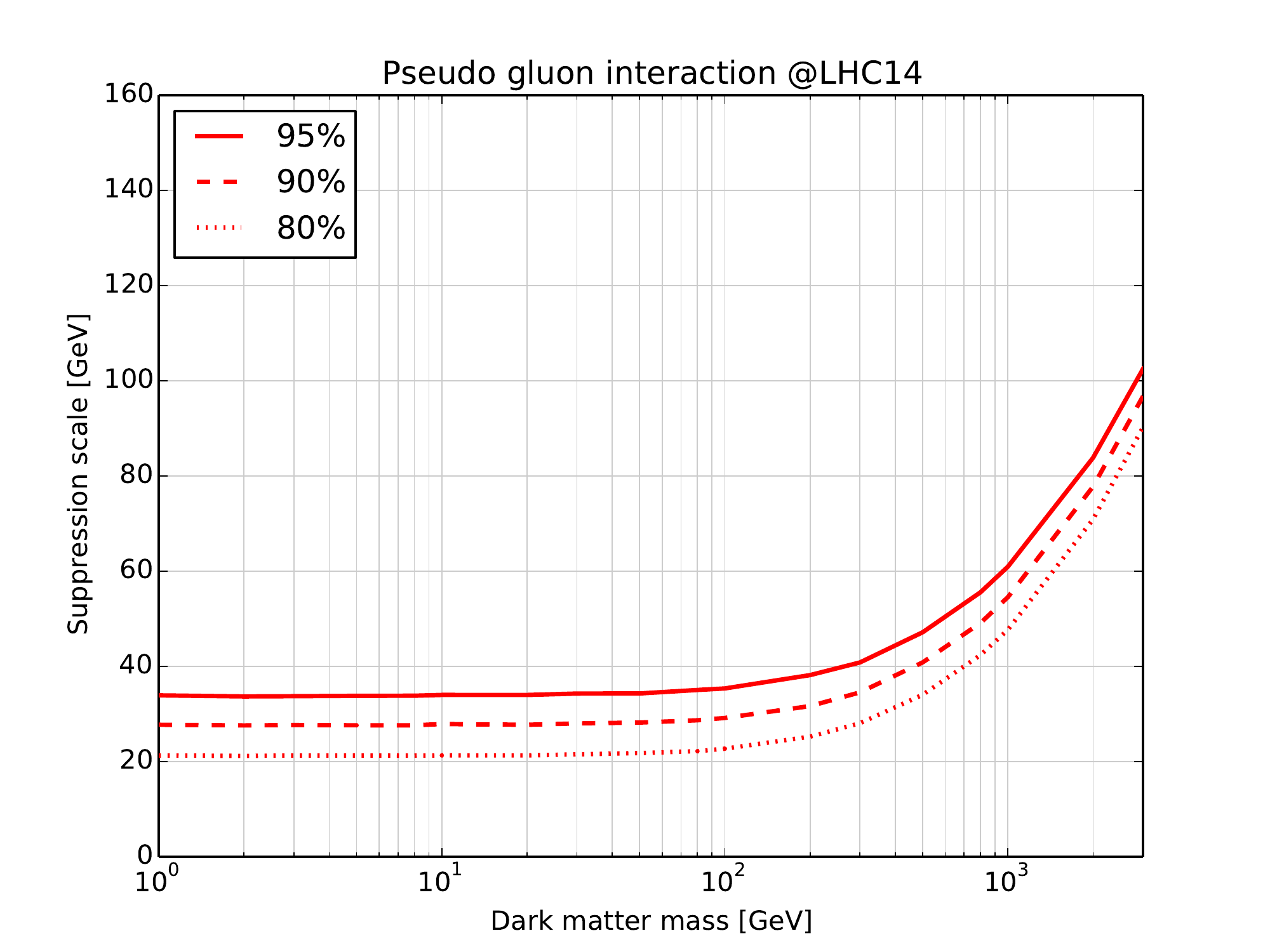}
\caption{The unitarity violation rate of the pseudo gluon interaction in LHC8(left) and LHC14(right). The meanings of the axes and lines are the same with Fig.~\ref{FigScalar}}
\label{FigGluon}
\end{figure}

Finally, the scalar dark matter which interact with gluon is studied~\footnotemark.
\footnotetext{
We have used the following formulae of the running QCD coupling,
\begin{align}
  \frac{g_S(Q)^2}{4\pi} = \frac{\al_S^Z}{1+\al_S^Z \frac{33-2N_f}{12\pi}\ln\frac{Q^2}{M_Z^2}},
\end{align}
where $\al_S^Z$ is the QCD $\al_S$ at the $Z$ boson peak and $N_f$ is the number of the quark flavor.
They are respectively $0.1184$ and $5$ in this paper.
The scale $Q^2$ has been chosen as the invariant mass of the dark matter pair for each event.
Even if we use the geometric means of $\sqrt{m_{DM}^2 +p_T^2}$ respecting the definition of MadGraph, differences are numerically negligible.
}
Since the gluon distribution in proton is much abundant in the region of the small energy fraction, the red/light grey lines are relatively closer than those of the vector interaction, nevertheless, this operator is also dimension six, see Fig.~\ref{FigGluon}.

For the gluon interaction, 90\% of events satisfy the unitarity condition~\eqref{EqGluon} if the dark matter mass is 800GeV, which is occasionally the same with the others. 
At that dark matter mass, the given bound of the suppression scale is 38GeV~\footnotemark.
\footnotetext{
 If the operator is defined without the loop factor and the symmetric factor, the bound is about 960GeV.
}

For proton-proton collision at 14TeV, the number of events violating the unitarity rapidly grows in the TeV region such as the vector interaction.

\section{Conclusion}
\label{SecConclusion}
Effective field theory has been used well in the dark matter search of LHC.
Since we do not know detail properties of the dark matter, this model independent way is quite useful to see strength of experimental bounds and compare results of other experiments looking for the dark matter.
However, in hadron colliders, collision energy of some events could be very large because of the compositeness of proton.
Cross sections of higher dimensional operators become large and larger as increasing of the collision energy.
Therefore, the S-matrix unitarity is suffered in the effective theory analyses.

In this paper, we have investigated the compatibility of the collider studies on the effective description of interaction between the complex scalar dark matter and the SM colored particles.
Concerning restrictions where 90\% of events satisfy the unitarity, studied three interactions are not valid if the dark matter is heavier than 800GeV in LHC8.
We have also pointed out the importance of operator dimension.
The 90\% constraints to the suppression scales on LHC8 occasionally coincide among the three interactions.
However, the unitarity condition for dimension six operators become much severe in LHC14, in spite that it becomes only a bit stronger for the dimension five operator.

Given results state that, using the effective field theory, collider searches of the dark matter is not valid if its mass is heavier than several hundreds GeV.
For spin independent interactions, the region is covered by direct detection experiments.
On the other hand, it is found that experimental constraints are weakened there for spin dependent interactions.
\section*{Acknowledgement}
The author thanks M. Endo for the discussion in the early stage of this work.
This work has been supported in part by the Ministry of Economy and Competitiveness (MINECO), grant FPA2010-17915, and by the Junta de Andaluc\'ia, grants FQM 101 and FQM 6552.

\appendix
\section{General formulae of the unitarity bound}
\label{AppGeneral}
We show the conditions of the $S$-matrix unitarity based on more general effective Lagrangian of the complex scalar dark matter and colored particles,
\begin{align}
 \mcl{L}_\text{eff}
 =&
   C_S \ph^\dag \ph (\bar{q} q)
 +iC_P \ph^\dag \ph (\bar{q} \ga_5 q)
 +iC_V (\ph^\dag \delfb_\mu \ph) (\bar{q} \ga^\mu q)
 +iC_A (\ph^\dag \delfb_\mu \ph) (\bar{q} \ga^\mu \ga_5 q) \n &
 +\frac{g_s^2 C_K }{(8\pi)^2} \ph^\dag \ph G^{a\mu\nu} G^a{}_{\mu\nu}
 +\frac{g_s^2 C_{CS}}{(8\pi)^2} \ph^\dag \ph G^{a\mu\nu} \tilde{G}^a{}_{\mu\nu},
\end{align}
where the Wilson coefficients are expressed such as $C^{\cdots}_{\cdots}$ instead of explicitly using the suppression scales.
With the discussion in Ref.~\cite{RefUnitarity}, the unitarity conditions are
\begin{align}
 \left( C_S^2 +C_P^2 \right) \frac{s\sqrt{1-4m_\text{DM}^2 /s}}{64\pi^2} < 1, \\
 \left( C_V^2 +C_A^2 \right) \frac{s^2(1-4m_\text{DM}^2 /s)^{3/2}}{288\pi^2} < 1,\\
 \left( C_K^2 +4C_{CS}^2 \right) \frac{g_s^4 s^2\sqrt{1-4m_\text{DM}^2 /s}}{65536\pi^6} < 1.
\end{align}
Since quark spins cannot be identified, conditions are averaged for initial states.

In terms of the chiral base, the conditions are written as,
\begin{align}
 C^S C^{S\ast} \frac{s\sqrt{1-4m_\text{DM}^2 /s}}{64\pi^2} < 1, \\
 \left( C^{L2} +C^{R2} \right) \frac{s^2(1-4m_\text{DM}^2 /s)^{3/2}}{576\pi^2} < 1,
\end{align}
where the coefficients are defined as follows,
\begin{align}
 C^S =& C_S +iC_P, \\
 C^L =& C_V -C_A,\qquad
 C^R =  C_V +C_A.
\end{align}

The above inequalities are easily derived for the real scalar dark matter.
We consider the following effective Lagrangian,
\begin{align}
 \mcl{L}_\text{eff}
 =&
   \frac{C'_S}{2} \ph'^2 (\bar{q} q)
 +i\frac{C'_P}{2} \ph'^2 (\bar{q} \ga_5 q)
 +\frac{g_s^2 C'_K }{2(8\pi)^2} \ph'^2 G^{a\mu\nu} G^a{}_{\mu\nu}
 +\frac{g_s^2 C'_{CS}}{2(8\pi)^2} \ph'^2 G^{a\mu\nu} \tilde{G}^a{}_{\mu\nu},
\end{align}
where $\ph'$ is the real scalar dark matter.
The perturbative unitarity conditions are
\begin{align}
 \left( C_S^{\prime 2} +C_P^{\prime 2} \right) \frac{s\sqrt{1-4m_\text{DM}^2/s}}{64\pi^2} < 2, \\
 \left( C_K^{\prime 2} +4C_{CS}^{\prime 2} \right) \frac{g_s^4 s^2\sqrt{1-4m_\text{DM}^2/s}}{65536\pi^6} < 2.
\end{align}
Expressions with the chiral base are trivial.


\end{document}